\documentclass[preprint,prlshowpacs,preprintnumbers,amsmath,amssymb,showkeys]{revtex4}

\usepackage{graphicx}
\usepackage{dcolumn}
\usepackage{bm}

\begin{document}


\title{
Dzyaloshinsky-Moriya interaction and long life time of the spin state in the Cu$_3$ triangular spin cluster 
by inelastic neutron scattering measurements
}

\author{Kazuki Iida$^1$}
 \email{iida2@issp.u-tokyo.ac.jp}
\author{Yiming Qiu$^{2,3}$}
\author{Taku J Sato$^1$}
\affiliation{$^1$Neutron Science Laboratory, Institute for Solid State Physics, University of Tokyo, Kashiwa 5-1-5, Chiba 277-8581, Japan}
\affiliation{$^2$NIST Center for Neutron Research, National Institute of Standards and Technology, Gaithersburg, Maryland 20899, USA}
\affiliation{$^3$Department of Materials Science and Engineering, University of Maryland, College Park, Maryland 20742, USA}
\date{\today}

\begin{abstract}
Inelastic neutron scattering (INS) experiments have been performed 
on the Cu$_3$ triangular molecular nanomagnet using powder samples.
In the medium resolution INS spectrum measured, there are two peaks at $\hbar\omega=0.5$ and 0.6~meV.
Comparing the observed $Q$ dependences of these peaks with calculations, 
these two INS peaks originate from the Cu$_3$ cluster.
From the observed peak position, width, and intensity, we have determined the optimum parameters of the spin Hamiltonian 
consisted of the antiferromagnetic exchange and the Dzyaloshinsky-Moriya interactions, 
which can also reproduce the magnetic susceptibility measurement.
In addition, we have directly observed that the ground state quartet is split into two doublets with the energy separation 
of 0.103~meV using high-resolution neutron spectroscopy, 
which exactly corresponds to that expected from the optimum parameters obtained from the medium resolution experiment.
The temperature dependences of the integrated intensities of the 0.5 and 0.6~meV peaks are well reproduced 
by the Boltzmann distribution of the energy levels of the model Hamiltonian below 10~K.
Furthermore, the inelastic peaks were visible even at very high temperatures as 50~K.
This indicates extraordinary weak coupling between phonons (or any other perturbations) and spin states 
in the Cu$_3$ cluster, compared to the other known molecular nanomagnets.
\end{abstract}

\pacs{Valid PACS appear here}


\maketitle

\section{Introduction}
Quantum phenomena in magnetism sometimes appear in the macroscopic measurement.
Especially, molecular nanomagnets~\cite{MolecularNanomagnet} have been providing rich playgrounds to investigate new types of such quantum phenomena.
For instance, Mn$_{12}$~\cite{Mn12} and Fe$_8$~\cite{Fe8}, which are the most widely investigated spin clusters realized as the single molecular magnet, 
show a superparamagneticlike behavior and a quantum tunneling between the total spin states.
Cr$_7$Ni~\cite{Cr7Ni}, where one Cr ion ($S=3/2$) of the Cr$_8$-ring~\cite{Cr8} is substituted by a Ni ion ($S=1$), 
is also a good example, showing a quantum coherence between the total spin states at the level-crossing field.

\begin{figure}[b]
\includegraphics[width=8.54cm, height=4.2cm]{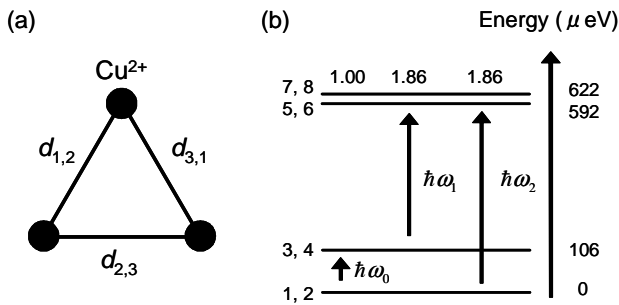}
\caption{\label{Fig:Structure}
(a) A structure of the Cu$_3$ spin cluster, where each circle represents the Cu$^{2+}$ ion.
The distances between Cu$^{2+}$ ions $d_{1,2}$, $d_{2,3}$, and $d_{3,1}$ are written in the text.
(b) Schematic view of the energy levels of the model Hamiltonian obtained by the optimum parameters given in Eq.~(\ref{Eq:Parameters}); 
all lines stand for each doublets and the number next to the lines is the label of the energy levels.
$\hbar\omega_0$ represents the splitting of the $S_\text{total}=1/2$ quartet by the DM interaction, 
whereas $\hbar\omega_1$ and $\hbar\omega_2$ correspond to the INS peaks at 0.5 and 0.6~meV in the INS spectra measured at HER 
as shown in Figs.~\ref{Fig:INS} and \ref{Fig:INS_Cal}.
The values ahead of arrows are the ratio of the calculated INS intensities of each excitations at $|Q|=0.95$~$\text{\AA}$$^{-1}$ divided by that of $\hbar\omega_0$.
}
\end{figure}

Another interesting issue is a magnetization reversal with a pronounced hysteresis observed in V$_{15}$~\cite{V15_1}.
This phenomenon can be quantitatively explained by the dissipative two-level model 
consisted of a Landau-Zener-St\"{u}ckelberg (LZS) transition and a phonon bottleneck effect.
For applying the LZS transition, the Dzyaloshinsky-Moriya (DM) interaction~\cite{DM_1,DM_2} (or the hyperfine interaction~\cite{HyperFine}) performs an important role 
to realize the avoided level crossing of the $S_\text{total}=1/2$ ground state doublets at zero field, 
where $S_\text{total}$ represents the total spin of the cluster in this paper.
Indeed, the zero field splitting of the ground state, which is a direct consequence of the DM interaction, 
was observed by the inelastic neutron scattering (INS) measurement~\cite{V15_2}.
In addition, from the result of changing the thermal coupling to the reservoir in the magnetization measurement, 
the phonon bottleneck effect is revealed to be the origin of the hysteresis~\cite{V15_3}.
NaFe$_6$~\cite{NaFe6} is another example in this category, 
which shows hysteresis loop near the level crossing field in the magnetic torque curve caused by the phonon bottleneck effect.
On the other hand, the long life time of the spin state is the origin of the hysteresis in the V$_6$ cluster~\cite{V6}, 
which also shows the magnetization reversal with hysteresis.

Recently, it has been found that the spin trimer clusters such as V$_3$~\cite{V3_1}, Cu$_3$As~\cite{Cu3As_1}, and Cu$_3$Sb~\cite{Cu3Sb_2,Cu3Sb_1} 
show a similar but more intriguing phenomenon, which is a half-step magnetization change with a milli-second order hysteresis.
The DM interaction~\cite{Cu3As_1,DM_1} is again the key to realize the avoided level crossing of the total spin, 
when the ground state changes to the $S_\text{total}=3/2$ state from the $S_\text{total}=1/2$ state.
The long life time of the spin state is inferred to cause the hysteresis in the triangular cluster system.
To date, there are only few reports~\cite{Cu3As_1,Cu3Sb_1} on the direct observation of the DM interaction, 
although we think the data are not sufficiently convincing.
In addition, none for the origin of the long life time for these spin trimer system.
These two facts are necessary to confirm the validity of the model for the half-step change with hysteresis in the pulsed-field magnetization curve.

Na$_{12}$[Cu$_3$(SbW$_{9}$O$_{33}$)$_2$ (H$_2$O)$_3$]$\cdot$46H$_2$O (hereafter, Cu$_3$ in short)~\cite{Synthesis_1} is one of such spin trimer clusters.
Several measurements such as magnetic susceptibility, magnetization in pulsed field, electric spin resonance, and nuclear magnetic resonance 
have been performed~\cite{Cu3Sb_1,Cu3Sb_2}.
Three Cu$^{2+}$ ions are placed at the distances of $d_{1,2}=d_{2,3}=4.871$ and $d_{3,1}=4.772$~$\text{\AA}$ as shown in Fig.~\ref{Fig:Structure}(a).
Cu$^{2+}$ ions with $S=1/2$ are coupled by the antiferromagnetic exchange interaction, as indicated by the magnetic susceptibility measurement.
The exchange path is thought to be Cu-O-W-O-W-O-Cu~\cite{Cu3Sb_2}.
There is no inversion symmetry at the center of any two Cu$^{2+}$ ions, suggesting the existence of the DM interaction in addition to the super exchange interactions.
Consequently, the following spin Hamiltonian has been proposed~\cite{Cu3Sb_1}:
\begin{eqnarray}
\mathcal{H}&=&\sum_{i=1}^{3}\sum_{\alpha}^{x,y,z}\left[-J^\alpha_{i,i+1}S_i^\alpha S_{i+1}^\alpha+\vec{D}_{i,i+1}\cdot\left(\vec{S}_i\times\vec{S}_{i+1}\right)\right]\nonumber\\
&+&\mu_\text{B}\sum_{i=1}^3\vec{S_i}\cdot\tilde{g}\cdot\vec{B},\label{Eq:Hamiltonian}
\end{eqnarray}
where $J_{i,i+1}^\alpha$ and $D_{i,i+1}^\alpha$ are the $\alpha$-component of the exchange and DM interactions between the $i$-th and $(i+1)$-th Cu$^{2+}$ ions, and 
$\mu_\text{B}$ is the Bohr magneton.
The reported energy level scheme consists of the ground state with $S_\text{total}=1/2$, 
the first excited state with $S_\text{total}=1/2$ at 100~$\mu$eV higher than the ground state, 
and two weakly split $S_\text{total}=3/2$ doublets at 580~$\mu$eV higher~\cite{Cu3Sb_1}.

The splitting of the $S_\text{total}=1/2$ quartet into two doublets is due to the DM interaction~\cite{notice_1}.
Thus, the DM interaction can be observed directly by detecting the excitation at $\hbar\omega_0$.
On the other hand, origin of the long life time of the spin state may be elucidated by directly observing the life time of the excitation levels, 
$i$.$e$., the broadening of the excitation peaks in INS spectra.
In this paper, we will discuss these features using neutron scattering spectroscopy.
First, we obtain the parameters of the model Hamiltonian of the Cu$_3$ spin cluster 
by observing the excitations to $S_\text{total}=3/2$ from $S_\text{total}=1/2$.
Then, we observe the DM interaction microscopically by detecting the splitting of the $S_\text{total}=1/2$ quartet.
Finally, we investigate the temperature dependence of the INS peaks to get an insight into the origin of the long life time of the spin state in Cu$_3$.

\section{Experimental Details}
The powder sample was prepared by the procedure reported in the Refs.~\onlinecite{Synthesis_1} and \onlinecite{Synthesis_2}, 
and the deuterated powder sample was also prepared for the high-energy-resolution INS measurement.
Magnetic susceptibility measurement of 45.3~mg non-deuterated powder sample was performed with a SQUID magnetometer in the temperature range of $1.8\le T\le300$~K.

A part of INS experiments with about 18.2~g non-deuterated powder sample was performed using the triple-axis spectrometer ISSP-HER, 
installed at the JRR-3 research reactor (Tokai, Japan). 
We have employed vertically focusing monochromator to select incident neutron wavelength, 
whereas double focusing ($i.e.$ both horizontal and vertical focusing) technique was used for the analyzer.
Pyrolytic graphite (PG) 002 reflections were used both for the monochromator and analyzer.
The spectrometer was operated in the fixed-final-energy mode with $E_\text{f}=2.4$~meV, 
resulting in the instrumental resolution of 61~$\mu$eV (FWHM, or full width at half maximum) at the elastic position.
The resolutions at $\hbar\omega=0.49$ and 0.60~meV are estimated as 68 and 71~$\mu$eV (FWHM), assuming the Cooper-Nathans type resolution function~\cite{ResolutionFunction_1}.
The higher-energy harmonics were eliminated using the cooled Be filter.
The non-deuterated powder sample was sealed in the aluminum sample can filled with the $^4$He exchange gas, 
which is set to a closed-cycle $^3$He refrigerator with the lowest working temperature of about 0.7~K.
An ILL Orange cryostat was also used in the measurement of $Q$ dependences as shown in Figs.~\ref{Fig:Q}(b) and \ref{Fig:Q}(d).

An INS experiment was also performed using the disk chopper time-of-flight spectrometer DCS installed 
at NIST Center for Neutron Research (Gaithersburg, USA) with $E_\text{i}=1.0$~meV.
The resolution at elastic position was obtained as 18.7~$\mu$eV (FWHM), and the resolutions at $\hbar\omega=-0.1$ and 0.1~meV were estimated as 22.6 and 15.1~$\mu$eV (FWHM), 
respectively~\cite{ResolutionFunction_2}.
The deuterated powder sample of about 4.7~g was put in the aluminum sample can, and set to the ILL Orange cryostat, with which the lowest working temperature is 1.5~K.

\section{Experimental Results}
\begin{figure*}[t]
\includegraphics[width=17.64cm, height=9.8cm]{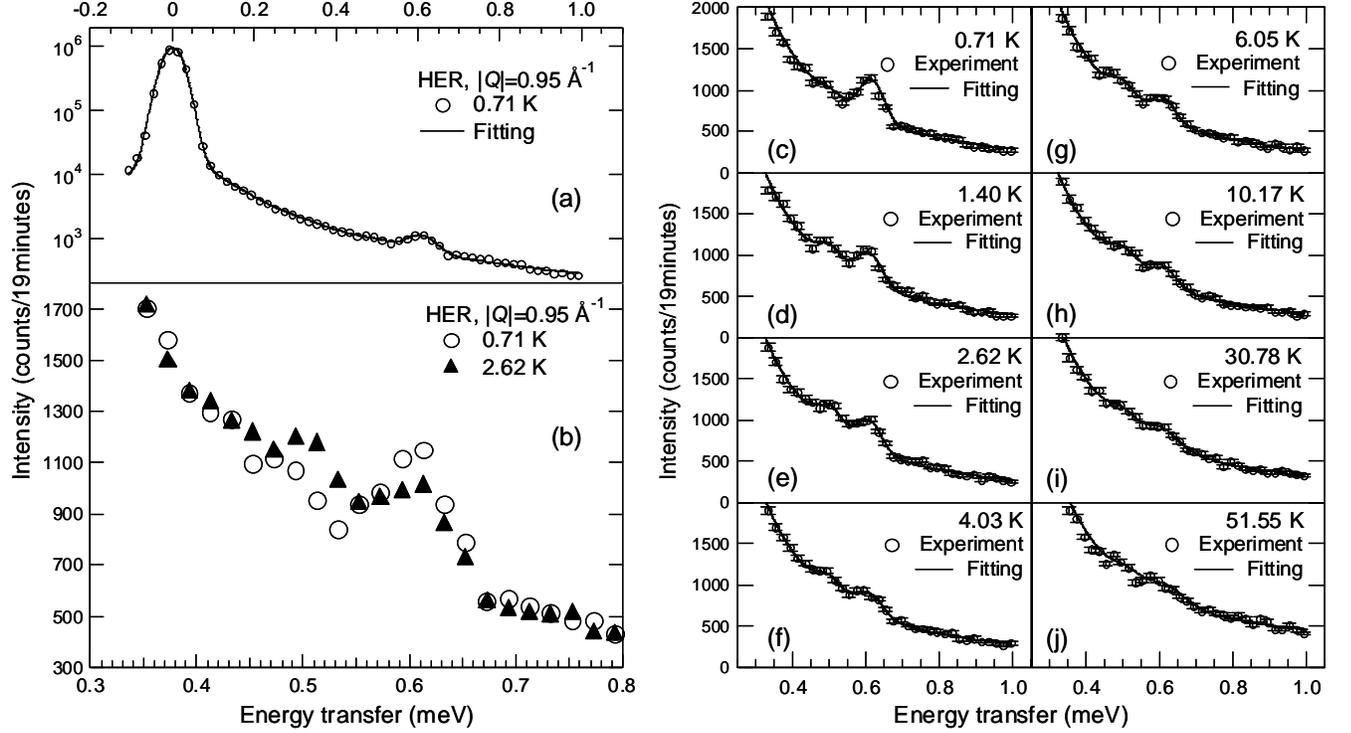}
\caption{\label{Fig:INS}
(a) The INS spectrum at $|Q|=0.95$~$\text{\AA}$$^{-1}$ and $T=0.71$~K measured at HER is shown.
Note the log scale for the vertical axis.
A solid line is the fitting result of Eqs.~(\ref{Eq:Fitting}), (\ref{Eq:Fitting_Inela}), and (\ref{Eq:Fitting_Incoh}).
(b) The INS spectra of $|Q|=0.95$~$\text{\AA}$$^{-1}$ at $T=0.71$ and 2.62~K.
The error bars of the observed data in both figures are within each marks, and 
the error bars in all figures represent one standard deviation.
The INS spectra of $|Q|=0.95$~$\text{\AA}$$^{-1}$ at (c) $T=0.71$~K, (d) 1.40~K, (e) 2.62~K, (f) 4.03~K, (g) 6.05~K, (h) 10.17~K, (i) 30.78~K, and (j) 51.55~K measured at HER 
and the fitting results of Eqs.~(\ref{Eq:Fitting}), (\ref{Eq:Fitting_Inela}), and (\ref{Eq:Fitting_Incoh}), are shown respectively.
}
\end{figure*}

An INS spectrum at $|Q|=0.95$~$\text{\AA}$$^{-1}$ and $T=0.71$~K measured at HER is shown in Fig.~\ref{Fig:INS}(a).
Although the strong incoherent scattering from hydrogen centered at the elastic position was observed as the background, 
there is an obvious INS peak around $\hbar\omega=0.6$~meV.
To focus on the INS peak, we plotted INS spectra at $|Q|=0.95$~$\text{\AA}$$^{-1}$, and $T=0.71$ and 2.62~K 
in the range of $0.3\le\hbar\omega\le 0.8$~meV in Fig.~\ref{Fig:INS}(b).
The spectrum at 0.71~K exhibits a weak peak at 0.5~meV as well as the pronounced peak at 0.6~meV.
At 2.62~K, the intensity at 0.5~meV increases, whereas the intensity of the 0.6~meV peak decreases.
The INS spectra of $|Q|=0.95$~$\text{\AA}$$^{-1}$ at various temperatures measured at HER are also shown in Figs.~\ref{Fig:INS}(c)$-$\ref{Fig:INS}(j).
The intensity of the 0.5~meV peak increases as temperature increases up to 2.62~K and decreases as temperature increases above 2.62~K, 
whereas the intensity of the 0.6~meV peak monotonically decreases as temperature increases.
As seen in Fig.~\ref{Fig:INS}, the magnetic signals are rather smeared by the strong incoherent background.
Therefore, to see the magnetic component clearly, the incoherent intensity was removed by the following procedure.
The INS spectrum is fitted by $I^\text{fit}$ defined as
\begin{equation}
I^\text{fit}(\hbar\omega)=I^\text{inela}(\hbar\omega)+I^\text{incoh}(\hbar\omega).\label{Eq:Fitting}
\end{equation}
$I^\text{inela}$ is the inelastic intensity and $I^\text{incoh}$ is the incoherent intensity defined as follows: 
\begin{eqnarray}
I^\text{inela}(\hbar\omega)&=&\sum_{i=1}^2c_i\frac{2\sqrt{\ln{2}}}{\sqrt{\pi}\Gamma_i}\label{Eq:Fitting_Inela}\\
&\times&\exp{\left[-\frac{4\ln{2}(\hbar\omega-\hbar\omega_i)^2}{\Gamma_i^2}\right]},\nonumber
\end{eqnarray}
\begin{eqnarray}
I^\text{incoh}(\hbar\omega)&=&c_3\frac{2\sqrt{\ln{2}}}{\sqrt{\pi}\Gamma_3}\exp{\left[-\frac{4\ln{2}(\hbar\omega)^2}{\Gamma_3^2}\right]}\label{Eq:Fitting_Incoh}\\
&+&c_4\frac{\Gamma_4}{\Gamma_4^2+(\hbar\omega)^2}\nonumber
\end{eqnarray}
where $c_i$ and $\Gamma_i$ represent an intensity normalization factor and a width of the Gaussian or Lorentzian function, 
whereas $\hbar\omega_i$ stands for a center of the INS peak.
By fitting $I^\text{fit}$ to the INS spectra of $|Q|=0.95$~$\text{\AA}$$^{-1}$ at various temperatures, 
we obtained peak positions as $\hbar\omega_1=0.498(2)$ and $\hbar\omega_2=0.607(1)$~meV, where $\hbar\omega_1$ and $\hbar\omega_2$ are defined in Fig.~\ref{Fig:Structure}(b).
In the fitting procedure, $\hbar\omega_1$ and $\hbar\omega_2$ are assumed to be global parameters, 
and all the spectra at different temperatures, $T=0.71$, 1.40, 2.62, 4.03, 6.05, 10.17, 30.78, and 51.55~K, are simultaneously fitted.
The reported result~\cite{Cu3Sb_1} expects that the excitations are observed at $\hbar\omega=0.484$ and 0.584~meV, which almost correspond to the present INS result.
This fact and the $Q$ dependences discussed later confirm that these excitations originate from the Cu$_3$ spin cluster.

\begin{figure}[t]
\includegraphics[width=8.4cm, height=11.2cm]{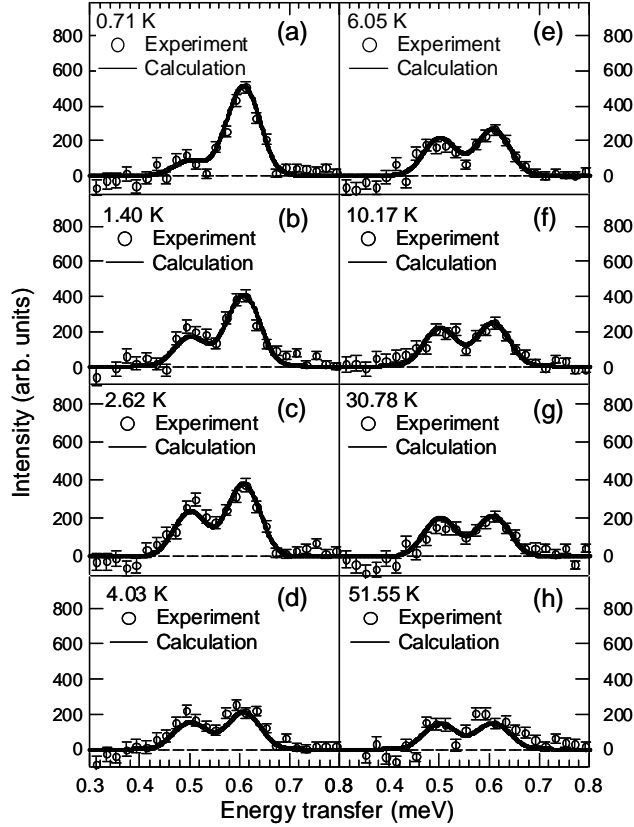}
\caption{\label{Fig:INS_Cal}
The experimental INS spectra measured at HER ($I^\text{exp}-I^\text{incoh}$) and the calculated intensities using Eq.~(3) in Ref.~\onlinecite{V3_2} of 
$|Q|=0.95$~$\text{\AA}$$^{-1}$ at (a) $T=0.71$~K, (b) 1.40~K, (c) 2.62~K, (d) 4.03~K, (e) 6.05~K, (f) 10.17~K, (g) 30.78~K, and (h) 51.55~K 
are illustrated, respectively.
}
\end{figure}

The fitting intensity $I^\text{fit}$ for $T=0.71$~K is shown by the solid lines in the Figs.~\ref{Fig:INS}(a) and \ref{Fig:INS}(c).
The INS spectrum is well fitted in the whole $\hbar\omega$ region.
The difference between the experimental intensity $I^\text{exp}$ and the incoherent intensity $I^\text{incoh}$ is plotted in Fig.~\ref{Fig:INS_Cal}(a).
In addition, we also showed the fitting intensities $I^\text{fit}$ compared with the experimental results at $T=1.40$, 2.62, 4.03, 6.05, 10.17, 30.78, and 51.55~K 
in Figs.~\ref{Fig:INS}(d)$-$\ref{Fig:INS}(j).
Noticeably, the INS spectra are also well fitted in the wide range of temperature.
The background subtracted spectra at these temperatures are plotted in Figs.~\ref{Fig:INS_Cal}(b)$-$\ref{Fig:INS_Cal}(h).
Temperature dependences of the peaks become clear; 
the intensities of the 0.5~meV peak at 1.40 and 2.62~K are bigger than that at 0.71~K and then decrease when temperature increases above 2.62~K, 
whereas the intensity of the 0.6~meV peak monotonically decreases with temperature increasing.
These temperature dependences can be explained by the Boltzmann distribution of the spin states as discussed later.

\begin{figure}[t]
\includegraphics[width=8.4cm, height=9.8cm]{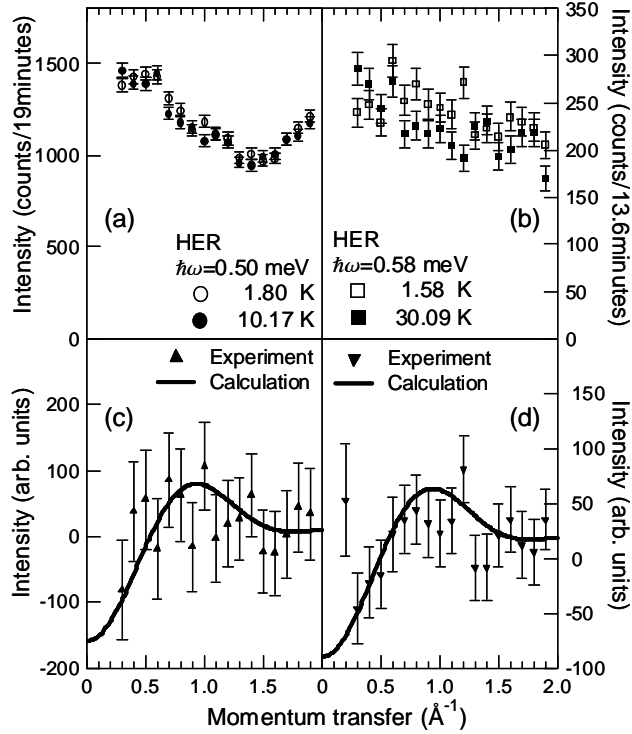}
\caption{\label{Fig:Q}
(a) The $Q$ dependences of the INS peak at $\hbar\omega=0.5$~meV, and $T=1.80$ and 10.17~K measured at HER.
(b) The $Q$ dependences of the INS peak at $\hbar\omega=0.58$~meV, and $T=1.58$ and 30.09~K measured at HER.
(c) The temperature difference of the $Q$ dependence at $\hbar\omega=0.5$~meV between 1.80 and 10.17~K and the calculated intensity.
(d) The temperature difference of the $Q$ dependence at $\hbar\omega=0.58$~meV between 1.58 and 30.09~K and the calculated intensity.
}
\end{figure}

The $Q$ dependences of $\hbar\omega=0.5$~meV at $T=1.80$ and 10.17~K measured at HER are shown in Fig.~\ref{Fig:Q}(a).
We also depicted the $Q$ dependences of $\hbar\omega=0.58$~meV at $T=1.58$ and 30.09~K in Fig.~\ref{Fig:Q}(b).
Both figures, at first glance, show only weak $Q$ dependences nor temperature dependences.
However, this is due to the strong incoherent contamination from hydrogen.
To clearly see the $Q$ dependence of the magnetic component, we subtracted the intensity at higher temperature from that at lower temperature.
The subtracted intensities are shown in Figs.~\ref{Fig:Q}(c) and \ref{Fig:Q}(d).
Both figures show decreasing behavior at low $Q$, the broad peaks around $|Q|=0.9$~$\text{\AA}$$^{-1}$, and mostly flat behavior at high $Q$.
They are indeed in good agreement with the geometry of the Cu$_3$ triangular spin cluster 
and the proposed interaction parameters between Cu$^{2+}$ ions as discussed later.

\section{Discussion}

In this section, we first determine the parameters in Eq.~(\ref{Eq:Hamiltonian}) from the excitations at $\hbar\omega=0.5$ and 0.6~meV 
corresponding to the excitations to $S_\text{total}=3/2$ from $S_\text{total}=1/2$.
Then, we show that the low-energy excitation at $\hbar\omega_0$, which is the splitting of the ground state quartet due to the DM interaction, 
is indeed observed by the high-energy-resolution neutron spectroscopy.
Finally, we discuss the temperature dependences of the INS peaks at $\hbar\omega=0.5$ and 0.6~meV 
to elucidate the origin of the long life time of the spin state in the Cu$_3$ cluster.

The procedure to calculate the neutron scattering function was reported in Ref.~\onlinecite{V3_2}, 
and we used the calculated intensity $I^\text{cal}(|Q|,\hbar\omega)$ as defined in Eq.~(3) in Ref.~\onlinecite{V3_2}.
We also used the magnetic form factor of Cu$^{2+}$ ions given in Ref.~\onlinecite{MFfactor}.
Since the peak widths in the fitting result are almost same as the resolution limited values below 10 K as shown in 
Fig.~\ref{Fig:Decoupling}(b) and  \ref{Fig:Decoupling}(d), and we only used the experimental data below 10 K to determine the optimum parameters in the spin Hamiltonian, 
we assume that the INS peaks have the resolution-limited widths in the present calculations for $I^\text{cal}(|Q|,\hbar\omega)$.
The temperature dependence will be discussed using the Boltzmann factor $p_\text{i}$ (used in Eq.~(2) in Ref.~\onlinecite{V3_2}) defined as follows:
\begin{equation}
p_\text{i}=\frac{e^{-\frac{E_\text{i}}{k_\text{B}T}}}{\sum\limits_{\text{j}=1}^8e^{-\frac{E_\text{j}}{k_\text{B}T}}}\label{Eq:p_i}
\end{equation}
where $k_\text{B}$ is the Boltzmann constant and the labels of the energy levels are defined in Fig.~\ref{Fig:Structure}(b).

\subsection{Parameters in the spin Hamiltonian}

To obtain the optimum parameters of the spin Hamiltonian, we performed least-squares fitting of $I^\text{cal}(|Q|,\hbar\omega)$ 
to the subtracted spectra ($I^\text{exp}-I^\text{incoh}$) at $T=0.71$, 1.40, 2.62, 4.03, 6.05, and 10.17~K as shown in Figs.~\ref{Fig:INS_Cal}(a)$-$\ref{Fig:INS_Cal}(f).
It may be noted that this time the scattering intensity of each peak is not an adjustable parameter, but only the overall intensity was optimized.
The calculated scattering intensities $I^\text{cal}(|Q|,\hbar\omega)$ for $T=0.71$, 1.40, 2.62, 4.03, 6.05, and 10.17~K 
are shown by the solid lines in Figs.~\ref{Fig:INS_Cal}(a)$-$\ref{Fig:INS_Cal}(f); a good match to the observation can be readily seen.
This satisfactory correspondence ensures the reliability of the estimated parameters.
The obtained optimum parameters are as follows:
\begin{eqnarray}
J_{1,2}^x=J_{1,2}^y                     &=& -4.19 \pm 0.03\ \text{K},\nonumber\\
J_{1,2}^z                               &=& -4.67 \pm 0.05\ \text{K},\nonumber\\
J_{2,3}^x=J_{2,3}^y=J_{3,1}^x=J_{3,1}^y &=& -4.14 \pm 0.01\ \text{K},\label{Eq:Parameters}\\
J_{2,3}^z=J_{3,1}^z                     &=& -4.42 \pm 0.02\ \text{K},\nonumber\\
D_{1,2}^z=D_{2,3}^z=D_{3,1}^z           &=& \ \ 0.66  \pm 0.01\ \text{K},\nonumber\\
D_{1,2}^x=D_{1,2}^y                     &=& \ \ 0.55  \pm 0.05\ \text{K}.\nonumber
\end{eqnarray}
The uncertainty ranges of the obtained parameters were estimated as the standard deviation of the Gaussian distribution using the linear approximation.
The energy levels using the optimum parameters are shown in Fig.~\ref{Fig:Structure}(b).
The $S_\text{total}=3/2$ states are almost degenerated whereas the $S_\text{total}=1/2$ quartet is split into two doublets.
The excitations, $\hbar\omega_0$, $\hbar\omega_1$, and $\hbar\omega_2$ are estimated as 0.106, 0.501, and 0.607~meV, respectively.
It should be noted that the optimum parameters are within the 10~\% difference of the reported parameters~\cite{Cu3Sb_1}, which are 
$J_{1,2}^x/\text{K}=-4.49$, $J_{1,2}^z/\text{K}=-4.54$, $J_{2,3}^x/\text{K}=-3.91$, $J_{2,3}^z/\text{K}=-3.96$, $D_{1,2}^z/\text{K}=0.517$, and $D_{1,2}^x/\text{K}=0.517$.
In addition, the estimated excitation energies of the reported result are almost the same as our result.

The $Q$ dependences at $\hbar\omega=0.5$ and 0.58~meV are also calculated using the optimum parameters; 
the solid lines in Figs.~\ref{Fig:Q}(c) and \ref{Fig:Q}(d) stand for the calculated intensities $I^\text{cal}(|Q|,\hbar\omega)$.
Both calculated results reproduce the temperature differences of the $Q$ scans well, 
which confirms that the 0.5 and 0.6~meV peaks are indeed consistent with the triangular geometry of the spin cluster and the interactions between the Cu$^{2+}$ spins.

\begin{figure}[t]
\includegraphics[width=8.4cm, height=4.9cm]{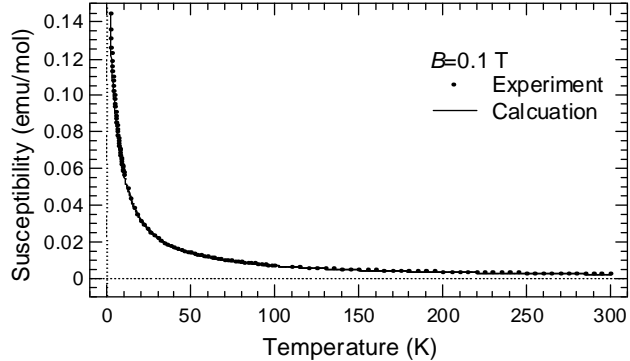}
\caption{\label{Fig:Susceptibility}
The experimental and calculated magnetic susceptibilities in $B=0.1$~T are shown.
}
\end{figure}

To further check the reliability of the model Hamiltonian, temperature dependence of the magnetic susceptibility is calculated using the optimum parameters.
In the calculation of the susceptibility, we use the reported value of the $g$ tensor~\cite{Cu3Sb_1} in Eq.~(\ref{Eq:Hamiltonian}).
Figure~\ref{Fig:Susceptibility} illustrates the comparison between observed magnetic susceptibility of the powder sample and the calculated susceptibility.
The observed susceptibility is well reproduced by the calculation in the wide temperature range, again confirming the validity of the present parameter estimation.
Therefore, the spin Hamiltonian Eq.~(\ref{Eq:Hamiltonian}) and its parameters Eq.~(\ref{Eq:Parameters}) 
can explain not only the INS results but also the susceptibility measurement.

\subsection{Ground state splitting due to the DM interaction}
\begin{figure}[t]
\includegraphics[width=8.4cm, height=5.6cm]{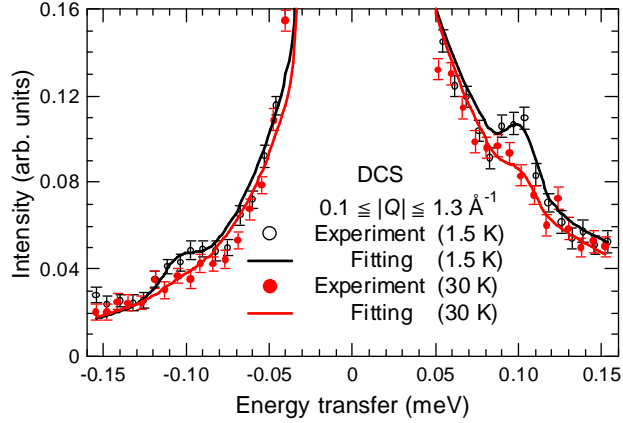}
\caption{\label{Fig:DCS}
(Color online)
The INS spectra of $0.1\le |Q|\le 1.3$~$\text{\AA}$$^{-1}$ at $T=1.5$ and 30~K measured at DCS using deuterated powder Cu$_3$ are plotted.
Both spectra are fitted by the same procedure in Fig.~\ref{Fig:INS} using Eqs.~(\ref{Eq:Fitting}), (\ref{Eq:Fitting_Inela}), and (\ref{Eq:Fitting_Incoh_2}).
}
\end{figure}

As shown in the above sections, the Cu$_3$ spin cluster most likely has energy level scheme summarized as Fig.~\ref{Fig:Structure}(b).
However, there is a possibility that the energy levels are miss-assigned; 
$i$.$e$., the DM interaction is very small and the anisotropy of the exchange interaction ($J_{i,i+1}^x\neq J_{i,i+1}^z$) is large, 
resulting in the large splitting of the $S_\text{total}=3/2$ quartet and the small splitting of the $S_\text{total}=1/2$ quartet.
We are now showing the definite evidence of the existence of the DM interaction in the model Hamiltonian 
by observing the excitation $\hbar\omega_0$, or the splitting between two $S_\text{total}=1/2$ doublets.
We note that the ground state splitting can not be possible by the other interactions than the DM interaction.
Figure~\ref{Fig:DCS} shows the INS spectra of deuterated powder Cu$_3$~\cite{notice_2}; 
$S(|Q|,\hbar\omega)$ at $T=1.5$ and 30~K measured at DCS is integrated in the range of $0.1\le |Q|\le 1.3$~$\text{\AA}$$^{-1}$.
There is surely an INS peak at $\hbar\omega=0.1$~meV in the $T=1.5$~K spectrum as expected from the HER result which is summarized in Fig.~\ref{Fig:Structure}(b).
To4 fit the INS spectra, we use different function for $I^\text{incoh}(\hbar\omega)$ in Eq.~(\ref{Eq:Fitting}), instead of Eq.~(\ref{Eq:Fitting_Incoh}), as follows:
\begin{eqnarray}
I^\text{incoh}(\hbar\omega)&=&c_3\frac{2\sqrt{\ln{2}}}{\sqrt{\pi}\Gamma_3}\exp{\left[-\frac{4\ln{2}(\hbar\omega)^2}{\Gamma_3^2}\right]}\label{Eq:Fitting_Incoh_2}\\
&+&c_4\frac{\Gamma_4}{\Gamma_4^2+(\hbar\omega)^2}\nonumber\\
&+&c_5\frac{\Gamma_5}{\Gamma_5^2+(\hbar\omega)^2}\times\frac{\hbar\omega}{1-\text{exp}\left(-\frac{\hbar\omega}{k_\text{B}T}\right)}.\nonumber
\end{eqnarray}
The difference of the shapes of the incoherent background (Eqs.~(\ref{Eq:Fitting_Incoh}) and (\ref{Eq:Fitting_Incoh_2})) 
may come from the different resolution functions of the spectrometers.
By fitting $I^\text{fit}$ of Eq.~(\ref{Eq:Fitting}) with Eqs.~(\ref{Eq:Fitting_Inela}) and (\ref{Eq:Fitting_Incoh_2}) to the INS spectra, 
the peak position is determined as $\hbar\omega=0.103(2)$~meV, which is almost exactly the same as the expected values, 
$\hbar\omega_0=0.102$~meV of the deuterated sample from the DCS result and $\hbar\omega_0=0.106$~meV of the non-deuterated sample from the HER result 
using the excitations $\hbar\omega_1$ and $\hbar\omega_2$.
The ground state splitting is surely confirmed by this high-energy-resolution experiment, and thus, 
we can conclude that the DM interaction surely exists in the Cu$_3$ spin cluster.
The splitting of the ground state quartet is of the similar magnitude as that in V$_3$~\cite{V3_2}.

\subsection{Spin and phonon states}
\begin{figure}[t]
\includegraphics[width=8.4cm, height=9.8cm]{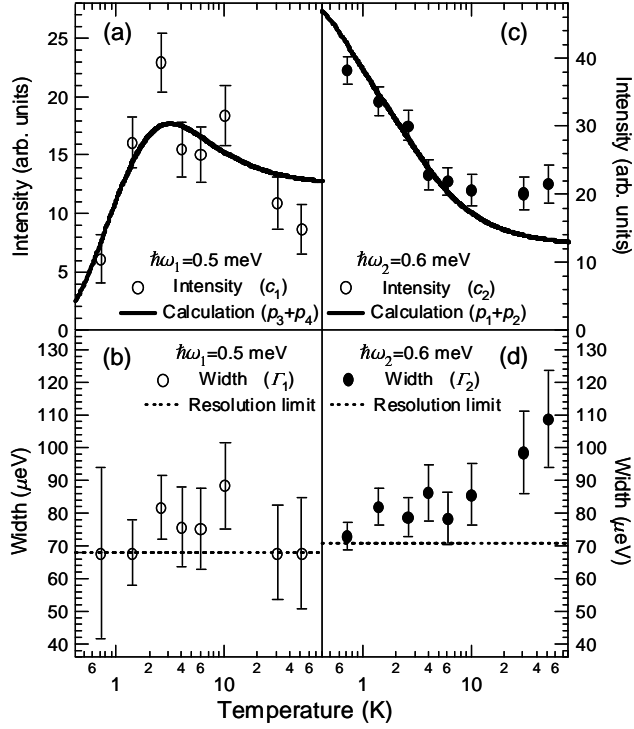}
\caption{\label{Fig:Decoupling}
(a) The integrated intensity of the peak at $\hbar\omega=0.5$~meV ($c_1$ in Eq.~(\ref{Eq:Fitting_Inela})) 
and the expected intensity ($(p_3+p_4)$ in Eq.~(\ref{Eq:p_i})).
We use a log scale for the horizontal axis here and subsequent panels.
(b) The width of the INS peak at $\hbar\omega=0.5$~meV ($\Gamma_1$ in Eq.~(\ref{Eq:Fitting_Inela})) and the resolution-limited width.
(c) The integrated intensity of the 0.6~meV peak ($c_2$) and the expected intensity ($p_1+p_2$).
(d) The width of the 0.6~meV peak ($\Gamma_2$) and the resolution-limited width.
}
\end{figure}

To discuss the temperature dependences of the INS peaks measured at HER in detail, the integrated intensities of the peaks at $\hbar\omega=0.5$ and 0.6~meV, $c_1$ and $c_2$ 
defined in Eq.~(\ref{Eq:Fitting_Inela}), are obtained from the fitting results as shown in Fig.~\ref{Fig:INS}~\cite{notice_3}.
The integrated intensities are shown in Figs.~\ref{Fig:Decoupling}(a) and \ref{Fig:Decoupling}(c), respectively.
Different temperature dependences are readily seen in the figures.
The 0.5~meV peak comes from the upper $S_\text{total}=1/2$ state to the $S_\text{total}=3/2$ states, 
whereas the 0.6~meV peak comes from lower $S_\text{total}=1/2$ to $S_\text{total}=3/2$ as shown in Fig.~\ref{Fig:Structure}(b).
The temperature dependences of the intensities of each peaks are governed by the Boltzmann factor of the initial state as described in Eq.~(2) in Ref.~\onlinecite{V3_2}.
Therefore, the peak intensity should follow the Boltzmann factor ($p_3+p_4$) or ($p_1+p_2$) as defined in Eq.~(\ref{Eq:p_i}) for the peak at 
$\hbar\omega_1=0.5$ or $\hbar\omega_2=0.6$~meV as far as other perturbations are small.

In Figs.~\ref{Fig:Decoupling}(a) and \ref{Fig:Decoupling}(c), the calculated intensities, $(p_3+p_4)$ and $(p_1+p_2)$, are compared with the experimental results.
The observed and calculated intensities for both INS peaks are in agreement below 10~K, whereas they are not above 10~K.
This feature is also confirmed by the temperature dependences of the peak widths of the INS peaks, $c_1$ and $c_2$ as defined in Eq.~(\ref{Eq:Fitting_Inela}).
Figures~\ref{Fig:Decoupling}(b) and \ref{Fig:Decoupling}(d) show that 
the temperature dependences of the peak widths of the INS peaks at $\hbar\omega=0.5$ and 0.6~meV, respectively.
Figure~\ref{Fig:Decoupling}(d) shows that the width of the 0.6~meV peak is much broader than the resolution limit value above 10~K, 
suggesting that the relaxation time of the spin state becomes shorter above 10~K.
Thus, the spin-phonon interaction may become relatively relevant above 10~K.
It should be noted that the spin-spin relaxation rate $1/T_2$ grew up under 10~K in the similar system Cu$_3$As~\cite{Cu3Sb_1}.
On the other hand, the calculated INS spectra for 30.78 and 51.55~K almost reproduce the experimental results 
in spite of such high temperature as shown in Figs.~\ref{Fig:INS_Cal}(g) and~\ref{Fig:INS_Cal}(h)~\cite{notice_4}.
In addition, the inelastic peaks were visible even at very high temperatures as 50~K.
These facts imply that the phonons are only perturbative to the spin states even above 10~K, only giving rise to the broadening effect.
We also note that the spin Hamiltonian Eq.~(\ref{Eq:Hamiltonian}) can reproduce the magnetic susceptibility up to 300~K very well as shown in Fig.~\ref{Fig:Susceptibility}.
Therefore, the spin state is only weakly influenced by the phonon (or other perturbations), and in particular, it is not disturbed below 10~K, 
at least in the neutron time scale.
It is very intriguing to study the origin of this decoupling between the phonon and spin states, and is left for future study.

\section{Conclusions}
INS experiments have been performed on the Cu$_3$ triangular spin cluster using both non-deuterated and deuterated powder samples.
Firstly, from the INS spectra measured at HER, we obtained the optimum parameters of the spin Hamiltonian listed in Eq.~(\ref{Eq:Parameters}).
These parameters and spin Hamiltonian can also reproduce the $Q$ dependences of the INS peaks as well as the magnetic susceptibility.
Secondly, we have directly observed the splitting of the ground state quartet due to the DM interaction at 0.1~meV; 
the value is almost exactly expected by the optimum parameters of the spin Hamiltonian determined using higher energy excitations.
Thirdly, the temperature dependences of the INS peaks at 0.5 and 0.6~meV 
suggest that the coupling between spin and phonon in Cu$_3$ is very weak, resulting in the rigid spin state, or the long life time of the spin state.
We, thus, conclude that they are the key features to explain the half-step magnetization change with the milli-second order hysteresis.

\section*{Acknowledgements}
One (K.I.) of us acknowledges for Global COE Program "the Physical Sciences Frontier", MEXT, Japan.
We also acknowledge for the financial support from the US-Japan Cooperative Program on Neutron Scattering.
Work at NCNR is in part supported by the National Science Foundation under Agreement No. DMR-0454672.
We would like to thank N. Aso, M. Yokoyama, T. Asami, Y. Kawamura, and J. R. D. Copley for their help in our INS experiments.


\begin{thebibliography}{30}
\bibitem{notice_1}
The site dependence of the exchange interaction ($i$.$e$., $J_{1,2}^x\neq J_{2,3}^x$) also divides the $S_\text{total}=1/2$ quartet.
But its contribution is small because Cu$_3$ is the approximately equatorial triangle, 
and indeed, our obtained parameters show small site dependence as written in Eq.~(\ref{Eq:Parameters}).
Thus, the gap, or the excitation $\hbar\omega_0$, mostly originates from the DM interaction.
\bibitem{notice_2}
From the INS spectrum of deuterated powder Cu$_3$ at $T=1.5$~K measured at DCS with $E_\text{i}=2.27$~meV (not shown), 
the parameters of the deuterated powder sample are determined by the excitations $\hbar\omega_1$ and $\hbar\omega_2$ as follows; 
$J_{1,2}^x/\text{K}=-4.30\pm0.06$, $J_{1,2}^z/\text{K}=-3.98\pm0.12$, $J_{2,3}^x/\text{K}=-4.18\pm0.03$, 
$J_{2,3}^z/\text{K}=-4.35\pm0.06$, $D_{1,2}^z/\text{K}=0.66\pm0.02$, and $D_{1,2}^x/\text{K}=0.37\pm0.14$.
The values are almost the same as the values of the non-deuterated powder sample (see Eq.~(\ref{Eq:Parameters})).
Thus, we can conclude that the deuterated Cu$_3$ spin cluster has almost same energy scheme as the non-deuterated Cu$_3$ spin cluster.
\bibitem{notice_3}
The number of excitations from $S_\text{total}=1/2$ to $S_\text{total}=3/2$ states are four, not two.
Then, we also fitted the INS spectra using four fixed excitations as follows; 
$\hbar\omega_1=0.486$, $\hbar\omega_2=0.516$, $\hbar\omega_3=0.592$, and $\hbar\omega_4=0.622$~meV in Eq.~(\ref{Eq:Fitting_Inela}).
But almost same temperature dependences of the integrated intensity and peak width as Fig.~\ref{Fig:Decoupling} are obtained.
Thus, we show the results using two excitations, $\hbar\omega_1$ and $\hbar\omega_2$ as defined in Fig.~\ref{Fig:Structure}(b), for simplicity.
\bibitem{notice_4}
We considered that the single INS peak has the resolution-limit width in the calculations for 30.78 and 51.55~K.
\end{thebibliography}

\end{document}